\documentclass[lettersize,journal]{IEEEtran}
\usepackage{amsmath,amsfonts}
\usepackage{algorithmic}
\usepackage{algorithm}
\usepackage{array}
\usepackage{subcaption}
\usepackage{textcomp}
\usepackage{stfloats}
\usepackage{url}
\usepackage{verbatim}
\usepackage{graphicx}
\usepackage{cite}
\usepackage{xcolor}
\makeatletter
\def\ps@IEEEtitlepagestyle{%
  \def\@oddfoot{\mycopyrightnotice}%
  \def\@evenfoot{}}
\def\mycopyrightnotice{%
  \hfill \footnotesize
  Accepted in: \textit{IEEE Communications Standards Magazine}, Oct. 23, 2025. 
  DOI: \texttt{10.1109/MCOMSTD.2025.3620005}\hfill
}
\makeatother
\begin{document}

\title{Energy-Efficient UAV-Mounted RIS for IoT: A Hybrid Energy Harvesting and DRL Approach}

\author{Mahmoud M. Salim, Khaled M. Rabie, \IEEEmembership{(Senior Member, IEEE)}, and Ali H. Muqaibel,
\IEEEmembership{(Senior Member, IEEE)}        % <-this % stops a space
\thanks{All authors are with the Center for Communication Systems and Sensing, King Fahd University of Petroleum and Minerals, Dhahran 31261, Saudi Arabia. Also, Khaled M. Rabie is affiliated with the Computer Engineering Department, and Ali H. Muqaibel is with the Electrical Engineering Department at King Fahd University of Petroleum and Minerals, Dhahran 31261, Saudi Arabia. Mahmoud M. Salim is the corresponding author (email: mahmoud.elemam@kfupm.edu.sa).}}
% <-this % stops a space
% \thanks{Mahmoud M. Salim is the corresponding author (email: mahmoud.elemam@kfupm.edu.sa). }}

\markboth{Journal of \LaTeX\ Class Files,~Vol.~14, No.~8, August~2021}%
{Shell \MakeLowercase{\textit{et al.}}: A Sample Article Using IEEEtran.cls for IEEE Journals}

\maketitle

\begin{abstract}
 % Integrating reconfigurable intelligent surfaces (RISs) with unmanned aerial vehicles (UAVs) introduces a flexible and energy-efficient solution for improving network coverage, spectral efficiency, and reliability, particularly in challenging environments.
 
 Many future Internet of Things (IoT) applications are expected to rely heavily on reconfigurable intelligent surface (RIS)-aided unmanned aerial vehicles (UAVs). However, the endurance of such systems is constrained by the limited onboard energy, where frequent recharging or battery replacements are required. This consequently disrupts continuous operation and may be impractical in disaster scenarios. To address this challenge, we explore a dual energy harvesting (EH) framework that integrates time-switching (TS), power-splitting (PS), and element-splitting (ES) EH protocols for radio frequency energy, along with solar energy as a renewable source. First, we present the proposed system architecture and EH operating protocols, introducing the proposed hybrid ES-TS-PS EH strategy to extend UAV-mounted RIS endurance. Next, we outline key application scenarios and the associated design challenges. After that, a deep reinforcement learning-based framework is introduced to maximize the EH efficiency by jointly optimizing UAV trajectory, RIS phase shifts, and EH strategies. The framework considers dual EH, hardware impairments, and channel state information imperfections to reflect real-world deployment conditions. The optimization problem is formulated as a Markov decision process and solved using an enhanced deep deterministic policy gradient algorithm, incorporating clipped double Q-learning and softmax-based Q-value estimation for improved stability and efficiency. The results demonstrate significant performance gains compared to the considered baseline approaches. Finally, possible challenges and open research directions are presented, highlighting the transformative potential of energy-efficient UAV-mounted RIS networks for IoT systems.

\end{abstract}

% Note that keywords are not normally used for peer review papers.
\IEEEpeerreviewmaketitle

% \textcolor{red}{More Ref. in the Intro}
% ******************************************************************
\section{Introduction}\label{Sec_Intro}

% \IEEEPARstart{S}{ixth} generation (6G) wireless networks are envisioned to revolutionize connectivity by enabling ubiquitous intelligence, extreme data rates, ultra-reliable low-latency communication, and sustainable energy efficiency \cite{wang-2023}. With data rates projected to reach $10–100$ Gbps on average and up to $1$ Tbps peak, together with precise positioning accuracies of $1$ cm indoors and $50$ cm outdoors, 6G must leverage intelligent infrastructure to meet these demands. Furthermore, the promise of $20$ year battery life for low-energy Internet of Things (IoT) devices requires changing energy-efficient and self-sustaining wireless environments \cite{nguyen-2021}.

\IEEEPARstart{I}{nternet} of Things (IoT) ecosystem is expected to undergo a significant transformation with the advent of sixth-generation (6G) wireless networks, which aim to support massive device connectivity, ultra-reliable low-latency communication, and sustainable energy operation \cite{wang-2023}. With anticipated data rates of 10–100 Gbps on average and peaks up to 1 Tbps, along with sub-meter-level positioning accuracy, 6G-enabled IoT networks must rely on intelligent infrastructure to meet the growing demands of diverse applications. Moreover, the vision of achieving a 20-year battery life for low-power IoT devices calls for a shift toward energy-efficient, self-sustaining wireless systems that can adapt to dynamic environmental and operational conditions \cite{keytech-ji-2021}.

At the heart of this transformation, reconfigurable intelligent surfaces (RIS) are redefining wireless communications by transforming static environments into programmable, adaptive elements \cite{shi-2022}. Composed of nearly-passive meta-material elements, RIS dynamically manipulates signal phase, amplitude, and polarization to optimize wireless propagation \cite{zhou-2023}. Unlike conventional relays that amplify signals at the cost of additional power consumption, RIS achieves similar or superior performance through passive beamforming, making it well-suited for large-scale, energy-efficient deployments. While static RIS are fixed to infrastructure, unmanned aerial vehicle (UAV)-mounted RIS offer greater flexibility and adaptability, maintaining both line-of-sight (LoS) and non-line-of-sight (NLoS) links in areas where terrestrial infrastructure is unavailable or obstructed \cite{deng-2024}. The authors in \cite{ris-satinoma-liu-2024} comprehensively reviewed recent advancements and standardization of UAV-mounted RIS, highlighting RIS's potential to enhance energy efficiency.

\begin{figure*}[ht]
\centering
\centerline{\includegraphics[width=2\columnwidth]{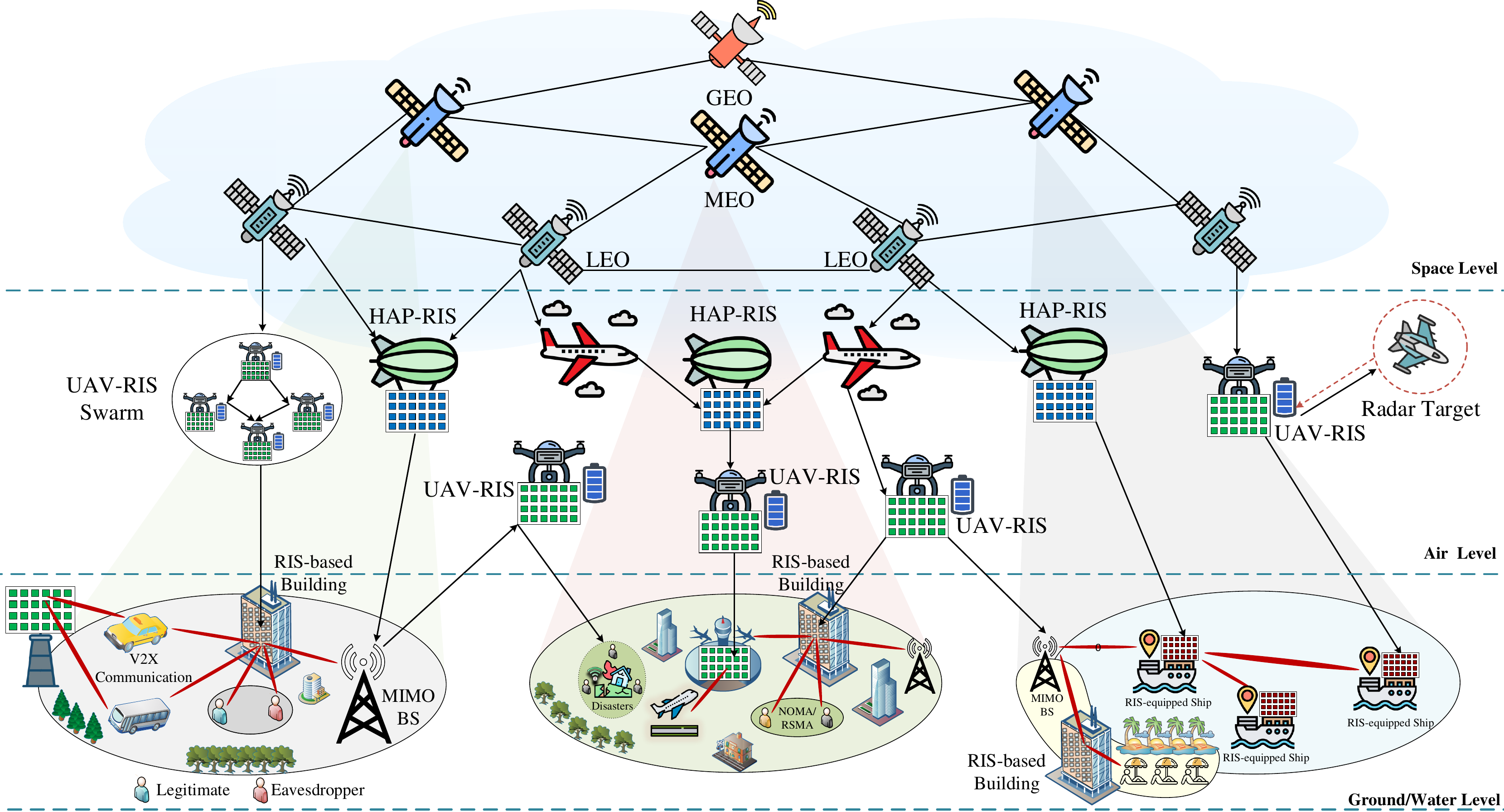}}
% \captionsetup{justification=centering}
\caption{Envisioned future 6G-enabled IoT networks with UAV-mounted RIS.}
\label{fig_Big_Pict}
\end{figure*}

As illustrated in Fig. \ref{fig_Big_Pict}, UAV-mounted RISs \cite{li-2020,ahmad-2022} function as dynamic air-layer nodes within space-air-ground integrated networks \cite{liu-2018}, seamlessly bridging spaceborne and terrestrial infrastructure to enhance connectivity in challenging environments. Operating across diverse scenarios, UAV-mounted RIS supports critical applications such as disaster recovery, vehicular networks, maritime communications, and aerial surveillance, ensuring reliable and adaptive wireless coverage. Furthermore, UAV-mounted RIS enhances multiple-input multiple-output (MIMO) diversity, improving spatial multiplexing and signal reliability. It also optimizes advanced multiple access schemes, such as rate-splitting multiple access (RSMA) and non-orthogonal multiple access (NOMA) \cite{ghosh-2024}, enabling more efficient spectrum utilization. Additionally, it reinforces network security by mitigating eavesdropping threats through adaptive signal propagation, making it an indispensable technology for next-generation wireless networks.
% **********************************************************************
% *********************************************************************
As discussed in \cite{peng-2023}, a major limitation of UAV-mounted RIS is their reliance on onboard batteries, which significantly impacts operational time and network sustainability. To address this challenge, research has explored radio frequency (RF) energy harvesting (EH), enabling UAV-RIS units to capture ambient RF signals \cite{peng-2023}. Beyond RF EH, solar energy serves as a complementary renewable source, offering a long-term sustainable solution \cite{salim-2022}. However, optimizing EH efficiency while maintaining reliable communication and mobility remains an open challenge.

Building on the need for energy-efficient UAV-mounted RIS networks, this paper proposes a dual EH framework integrating different RF EH protocols as well as solar energy to enhance UAV endurance in IoT systems. We introduce an intelligent optimization strategy leveraging deep reinforcement learning (DRL) to dynamically adjust RIS phase shifts, base station (BS) power allocation, and EH scheduling, ensuring efficient energy utilization and prolonged UAV operation. By jointly optimizing these parameters, we aim to achieve energy-efficient, self-sustaining UAV-mounted RIS networks capable of supporting future 6G-enabled IoT applications.

This discussion begins by presenting the UAV-mounted RIS system architecture and EH operating protocols in “System Architecture and Operating Protocols”. In this section, we introduce a hybrid approach that leverages spatial, time, and power dimensions to enhance UAV-RIS efficiency, ensuring optimal EH and communication performance in dynamic environments. The “Application Scenarios for UAV-Mounted RIS” section explores real-world use cases enabled by UAV-mounted RIS, highlighting their potential applications in physical-layer security, wireless coverage extension, disaster recovery, and smart city infrastructure. The “Design Challenges of UAV-Mounted RIS Networks” section examines key challenges, including energy and resource management, trajectory optimization, computational complexity, and practical impairments affecting UAV-RIS performance. To address these issues, we propose a DRL-based optimization algorithm in “Case Study: DRL-Based Optimization for Hybrid EH UAV-Mounted RIS Networks,” where we formulate the problem as a Markov decision process (MDP) and demonstrate the advantages of intelligent learning techniques. The “Numerical Analysis” section provides a performance evaluation of the proposed DRL-based algorithm through case studies comparing different EH protocols as well as other benchmarks. Finally, we discuss emerging research challenges and future directions in UAV-mounted RIS networks in “Future Directions in UAV-Mounted RIS Networks. Finally, the paper is concluded in the section “Conclusion.”
% *********************************************************************
\section{System Architecture and Operating Protocols} \label{Sec_SM}
% *********************************************************************
\subsection{System Architecture}
\begin{figure*}[t]
\centering
\centerline{\includegraphics[width=1.5\columnwidth]{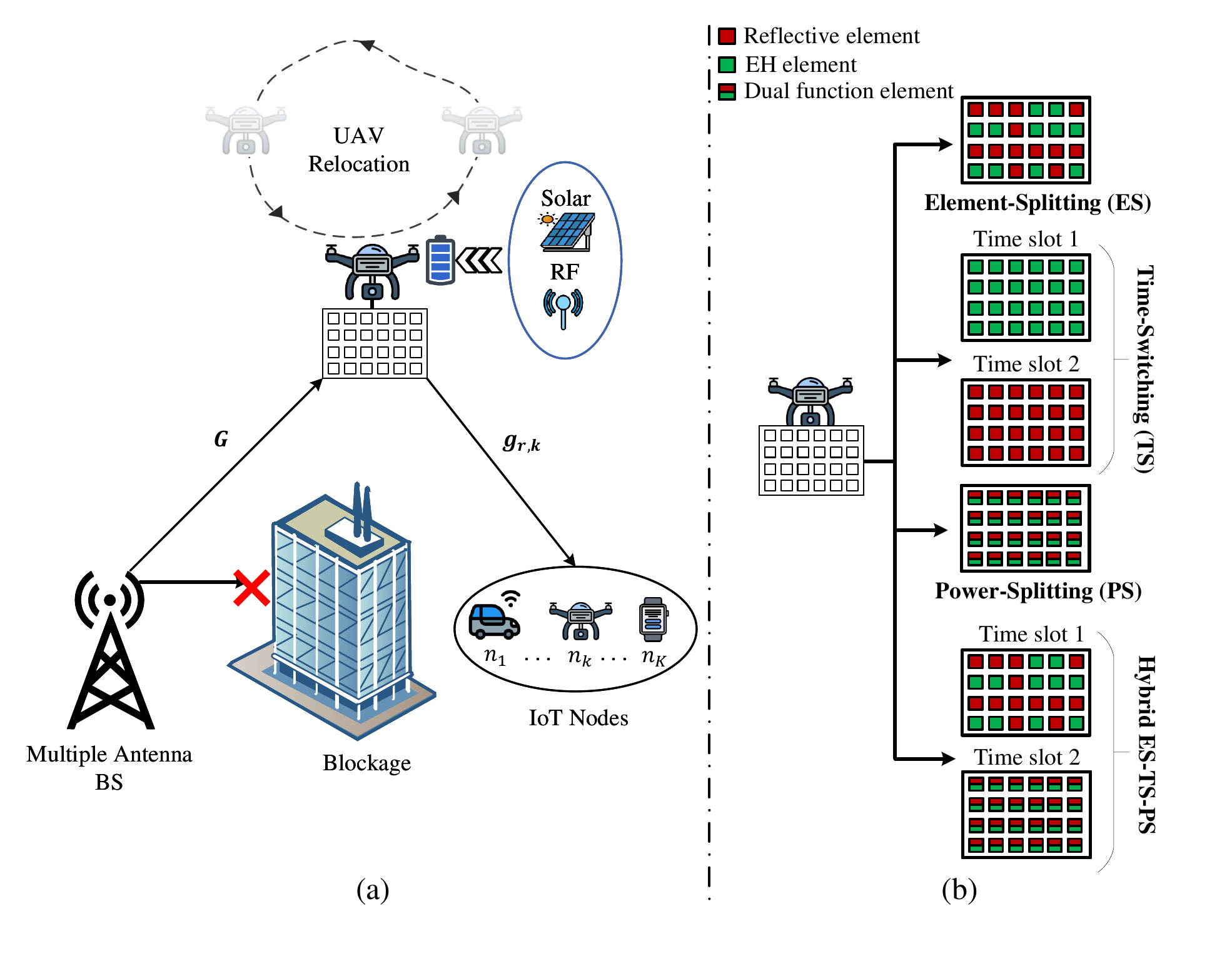}}
% \captionsetup{justification=centering}
\caption{Proposed UAV-mounted RIS framework (a) system architecture (b) RF EH operating protocols.}
\label{fig_SM}
\end{figure*}

% *********************************************************************
Fig.~\ref{fig_SM}(a) depicts the UAV-mounted RIS-assisted communication framework, where a multiple-antenna BS communicates with multiple mobile IoT nodes with single antennas via a UAV-mounted RIS unit. To address power constraints, the UAV-mounted RIS integrates RF EH to capture ambient RF signals and solar panels for sustainable energy, enhancing endurance and operational efficiency. The UAV-RIS unit employs a shared battery that follows a harvest-transmit-store mechanism.

The RIS elements within the UAV-mounted system are categorized into three functional types
\begin{itemize}
    \item \textbf{EH Elements:} These elements capture and convert ambient RF and solar energy into usable power, reducing the reliance on onboard batteries and prolonging the operational duration.
    \item \textbf{Reflective Elements:} These elements optimize passive beamforming to enhance LoS and NLoS links, improving IoT node spectral efficiency and interference mitigation without additional power consumption.
    \item \textbf{Dual-Function Elements:} These elements simultaneously perform EH and signal reflection, dynamically allocating energy between communication and power sustainability based on operating protocols.
\end{itemize}

A UAV-mounted RIS outperforms static RIS by integrating EH and mobility, optimizing power use and trajectory. Using RF and solar energy minimizes battery dependence while extending flight time. Through adaptive beamforming and energy management, communication and EH are efficiently balanced, ensuring optimal performance in dynamic environments.
% **********************************************************************
% *********************************************************************
% *********************************************************************
\subsection{Operating Protocols}
% *********************************************************************
To support both data transmission and UAV-RIS endurance, adaptive operating protocols adjust according to network conditions and energy availability. The UAV-mounted RIS operates primarily under element-splitting (ES), time-switching (TS), and power-splitting (PS) protocols. Besides, this work proposes a hybrid ES-TS-PS approach that dynamically allocates RIS elements across time slots and spatial regions, balancing power consumption, spectral efficiency, and resource allocation for flexible operation in dynamic environments. These protocols can also work with renewable energy (RE) through solar panels, further reducing the reliance on battery reserves on board.

The main idea behind these protocols is depicted in Fig.~\ref{fig_SM}(b) and can be presented as follows:
\begin{itemize}
    \item \textbf{ES Protocol}: The ES protocol statically assigns distinct groups of RIS elements to either signal reflection or EH, simplifying system design and ensuring stable operation. By dedicating each element to a specific task, this approach eliminates the need for continuous parameter adjustments, reducing computational complexity. However, in UAV-mounted RIS systems, this fixed allocation limits adaptability to real-time network variations. The mobility of UAVs introduces dynamic channel fluctuations, necessitating flexible resource allocation for optimal performance. The rigidity of the ES protocol might result in suboptimal efficiency, particularly in environments with frequent UAV movement and varying channel conditions. To address this, the element scheduling factor must be carefully optimized to balance EH and reflection under changing network conditions.
    \item \textbf{PS Protocol}: The PS protocol allows UAV-mounted RIS elements to simultaneously perform EH and signal reflection by dynamically splitting the received signal between the two functions. The proportion of received power allocated to EH and reflection is controlled by the PS factor, which must be carefully tuned to optimize both energy sustainability and communication performance. Unlike the ES protocol, which assigns roles to elements statically, PS provides adaptability to real-time network variations and UAV mobility. However, implementing PS introduces significant hardware complexity, requiring precise circuit design and advanced resource allocation strategies to optimize power splitting and ensure system efficiency.

    \item \textbf{Hybrid ES-TS-PS Protocol}: The proposed hybrid ES-TS-PS protocol integrates the strengths of ES, TS, and PS to optimize energy efficiency and communication performance across two time slots, regulated by the TS factor. In the first slot, a dynamic version of the ES protocol assigns RIS elements to EH and reflection roles based on an ES factor, which is optimized for each RIS element. In the second slot, the PS protocol allows elements to dynamically split received power between EH and reflection, governed by the PS factor. While this hybrid approach enhances efficiency, it also introduces significant computational complexity, necessitating precise synchronization of the TS factor and real-time optimization of ES and PS factors as well. Given the dynamic nature of network conditions and UAV mobility, intelligent algorithms are essential for efficient resource allocation, ensuring seamless mode transitions, optimal power distribution, and minimal system overhead.
    \end{itemize}

%********************************************************************* 
\section{Application Scenarios for UAV-Mounted RIS} 
%********************************************************************* 
\subsection{Physical Layer Security (PLS)} 
%********************************************************************* 
Ensuring secure communication in 6G-enabled IoT wireless environments is challenging, especially with eavesdroppers intercepting signals. UAV-mounted RIS enhances PLS in 6G-enabled IoT networks by dynamically adjusting trajectory and beamforming to direct signals toward legitimate IoT nodes while minimizing leakage. Artificial noise injection further degrades eavesdropper reception. Unlike static RIS, UAV-mounted RIS actively tracks and counters security threats while leveraging EH to sustain extended operations. By optimizing its energy efficiency, UAV-RIS ensures prolonged flight endurance, making it highly suitable for military, financial, and IoT applications requiring persistent security.
%*********************************************************************
\subsection{Public Safety and Disaster Response} 
%********************************************************************* 
In disasters where terrestrial infrastructure fails, 6G-enabled IoT systems powered by UAV-mounted RIS provide rapid-deployment connectivity for first responders and affected communities. By adjusting position and phase shifts it ensures low-latency communication between rescue teams and command centers. Unlike static networks, UAV-RIS in 6G-enabled IoT scenarios can relay mission-critical data and support real-time video streaming. Equipped with RF and solar EH, UAV-RIS maximizes power efficiency and extends operational endurance, ensuring continuous coverage during prolonged disaster relief efforts without frequent battery replacements or recharging.

%*********************************************************************
\subsection{Vehicle-to-Everything (V2X) Communication} 
%********************************************************************* 
Autonomous vehicles rely on low-latency, high-reliability communication for safe transport. UAV-mounted RIS in 6G-enabled IoT networks extends network coverage by optimizing beamforming and trajectory for seamless V2X communications. It mitigates urban signal blockages, optimizes NOMA and RSMA for spectral efficiency, and acts as an adaptive relay for fast-moving vehicles, ensuring real-time road hazard detection and platooning. UAV-RIS utilizes EH to minimize power dependency on conventional energy sources, enabling prolonged aerial support for V2X networks while maintaining energy-efficient operation.
%*********************************************************************
\subsection{UAV Swarm Coordination} 
%********************************************************************* 
Inter-UAV communication is vital for autonomous swarm operations in surveillance and environmental monitoring. UAV-mounted RIS enhances connectivity within 6G-enabled IoT networks by dynamically reflecting signals, minimizing interference, and maintaining strong LoS links. It supports cooperative sensing, synchronized navigation, and multi-UAV task execution while securing wireless links against eavesdropping and jamming. UAV-RIS optimizes its power allocation through EH from RF and renewable sources, reducing energy constraints and extending swarm endurance. This capability allows UAV swarms to operate efficiently in long-duration missions with minimal power depletion.

%*********************************************************************
\subsection{Remote and Maritime Connectivity} 
%********************************************************************* 
Providing reliable and energy-efficient connectivity in remote and maritime regions remains a significant challenge for 6G-enabled IoT systems due to limited infrastructure and high power constraints. UAV-mounted RIS serves as a flexible relay, ensuring seamless communication for offshore vessels, remote islands, and deep-sea research. By optimizing beamforming and positioning, it supports environmental monitoring, autonomous underwater vehicle coordination, and maritime logistics. Unlike satellites, UAV-RIS offers lower latency, adaptability, and extended endurance through RF and solar EH, reducing reliance on battery reserves. By adopting an energy-efficient approach, UAV-RIS optimally manages power consumption, ensuring prolonged operation and minimizing the need for frequent recharging or battery replacements. This energy-efficient operation allows IoT-driven UAV-RIS systems to sustain long-term communication coverage over vast oceanic regions without frequent power interruptions.
%*********************************************************************
\section{Design Challenges of UAV-Mounted RIS Networks}
%*********************************************************************
UAV-mounted RIS networks in 6G-enabled IoT environments introduce a complex interplay between energy efficiency, trajectory control, and resource optimization. Unlike static RIS, these systems operate in dynamic IoT scenarios, requiring continuous adaptation to changing channel conditions, power availability, and IoT node distribution. Addressing these challenges demands advanced strategies for EH, mobility management, and computational efficiency to ensure reliable, long-duration operation with minimal overhead. The following subsections outline key considerations in optimizing UAV-mounted RIS systems for 6G-enabled IoT networks.
%*********************************************************************
\subsection{Energy and Resource Management}
%*********************************************************************
Efficient energy and resource management is vital for UAV-mounted RIS networks operating in IoT environments to maintain continuous operation and reliable connectivity. Unlike terrestrial RIS, UAV-mounted RIS units should harvest energy from RF, solar, or hybrid sources to extend flight endurance. Optimized power allocation between communication, beamforming, and propulsion ensures sustained operation without degrading performance. UAV-mounted RIS should dynamically schedule EH and adjust beamforming based on real-time energy conditions and network demands. Intelligent resource allocation strategies enhance energy efficiency while preserving high-quality communication links.
%*********************************************************************
\subsection{Trajectory Optimization}
%*********************************************************************
UAV mobility necessitates continuous trajectory optimization to sustain reliable connectivity while minimizing energy consumption. Unlike fixed infrastructure, UAV-mounted RIS should dynamically adjust its flight path to optimize LoS and NLoS links. High-speed mobility introduces challenges such as Doppler effects, handover delays, and link instability, impacting network performance. Trajectory-aware beamforming mitigates these disruptions, enabling UAV-RIS to reposition based on real-time IoT node distribution, energy availability, and communication quality-of-service (QoS) requirements.
%*********************************************************************
\subsection{Practical Aspects and System Impairments}
%*********************************************************************
Several impairments impact UAV-mounted RIS networks in 6G-enabled IoT applications, affecting channel estimation, EH efficiency, and overall performance. Mobility-induced Doppler shifts, outdated channel state information (CSI) feedback, and estimation errors reduce beamforming accuracy, necessitating robust estimation techniques. Non-linear RF EH conversion and fluctuating incident energy levels require advanced power management to optimize EH and sustain UAV endurance. Hardware impairments, such as phase noise, amplifier distortions, and RIS element mismatches, further degrade performance. Given UAV-mounted RIS constraints in power and weight, real-time compensation algorithms and energy-aware beamforming strategies are crucial for mitigating impairments and ensuring robust operation in dynamic conditions.

%*********************************************************************
\subsection{Computational Complexity and Optimization Challenges}
%*********************************************************************
The joint optimization of UAV trajectory, RIS beamforming, power allocation, and EH scheduling presents a high-dimensional problem requiring intelligent real-time decision-making. Traditional methods, such as alternating optimization and convex approximations, offer structured solutions but suffer from high computational costs. Heuristic techniques reduce complexity but may yield suboptimal performance under dynamic conditions. DRL-based approaches, such as deep deterministic policy gradient (DDPG) algorithms, have emerged as promising solutions, allowing UAV-mounted RIS to efficiently adapt key parameters in response to time-varying environments. These AI-driven methods enhance decision-making while reducing computational overhead, making them well-suited for large-scale UAV-assisted RIS networks.
%*********************************************************************
\section{Case Study: DRL-Based Optimization for Hybrid EH UAV-Mounted RIS Networks}
%*********************************************************************
\subsection{System Model Description and Problem Formulation}
%*********************************************************************
To address the aforementioned design challenges in UAV-mounted RIS networks, we propose a DRL-based optimization framework for the hybrid EH-enabled UAV-mounted RIS communication systems. As illustrated in Fig. \ref{fig_SM}(a), we consider a communication system where a BS with \( Z \) antennas communicates with \( K \) single-antenna mobile IoT nodes via a UAV-mounted RIS with \( L \) reflective elements. The direct link between the BS and IoT nodes is assumed to be obstructed. The communication period is divided into \( T \) equal time slots, during which the UAV-mounted RIS harvests energy from incoming RF signals. In this regard, we adopt nonlinear models for TS, PS, and the proposed hybrid ES-TS-PS protocols, evaluating their performance separately to assess their impact on UAV-RIS EH efficiency. Additionally, the UAV-RIS unit collects solar energy at the beginning of each time slot. 

To capture real-world limitations, we incorporate CSI imperfections, introducing estimation errors and hardware impairments that cause distortion noise between the transmitter and receiver. Our objective is to develop a robust DRL-based optimization framework that maximizes EH efficiency and extends UAV-RIS endurance by jointly optimizing BS transmit power, RIS phase shifts, and EH protocol parameters (ES, TS, PS, and hybrid factors). The proposed framework ensures QoS requirements, power constraints, UAV mobility adaptation, and EH considerations, including causality and battery overflow constraints, making it an advanced solution for endurance-critical UAV-mounted RIS networks.

%*********************************************************************
\subsection{Markov Decision Process and DRL}
%*********************************************************************
To enable real-time adaptation and intelligent resource allocation, the UAV-mounted RIS network is formulated as an MDP, allowing it to learn from its environment and optimize key parameters dynamically. The MDP framework provides a structured approach where the system interacts with the environment, evaluates its performance, and refines its decisions over time through reinforcement learning.

The MDP formulation consists of the following components:
\begin{itemize}
    \item \textbf{State Space:} The state space comprises key system parameters, including the channel from the BS to the UAV-RIS, represented by the small-scale fading matrix \( \mathbf{G}_1 \), following a Rayleigh fading distribution. The channel between RIS element \( r \) and IoT node \( k \), denoted as \( \mathbf{g}_{r,k} \), follows a Rician fading distribution. Additionally, the state includes the positions of RIS elements \( \mathcal{P}^r_{i,j} \), IoT node antenna positions \( \mathcal{P}^k \), and the harvested RE energy \( \varepsilon^{RE} \). Furthermore, the state contains the action of the previous time slot $a^{(t-1)}$. These parameters collectively define the UAV-RIS system's current state and influence its decision-making process. The system state is expressed as 
    % {\scriptsize
    % \begin{equation}
    % s^{(t)}= \big[ \Re\{\mathbf{G}_1\}, \Im\{\mathbf{G}_1\}, \Re\{\mathbf{g}_{r,k}\}, \Im \{ \mathbf{g}_{r,k}\}, \mathcal{P}^r_{i,j}, \mathcal{P}^k, \varepsilon^{RE}, a^{(t-1)} \big], \nonumber
    % \end{equation}}
    \begin{align}
    s^{(t)} = \big[\, & \Re\{\mathbf{G}_1\},\ \Im\{\mathbf{G}_1\},\ \Re\{\mathbf{g}_{r,k}\},\ \Im\{\mathbf{g}_{r,k}\}, \nonumber \\
                     & \mathcal{P}^r_{i,j},\ \mathcal{P}^k,\ \varepsilon^{RE},\ a^{(t-1)} \,\big],
    \end{align}
    where \( \Re \{\cdot\} \) and \( \Im \{\cdot\} \) denote the real and imaginary components.
    \item \textbf{Action Space:} The actions involve adjusting the EH factors, including $\tau$ for TS, $\rho$ for PS, and $\tau$, $\rho$, and $\omega$ for the hybrid approach, along with configuring the RIS phase shifts $\theta_l$ and allocating BS transmission power $p_k$ to each IoT node $k$. Thus, the system action is defined as $a^{(t)} = \big[ \alpha, \theta_l, p_k \big], ~ \alpha \in \{ \tau, \rho \}$ when using the TS or PS protocols, and $a^{(t)} = \big[ \tau, \rho, \omega, \theta_l, p_k \big]$ when using the hybrid ES-TS-PS.
    \item \textbf{Reward Function:} The reward function is formulated to maximize EH efficiency while ensuring QoS, power constraints, and EH feasibility. The EH efficiency is defined as the ratio of the total harvested power to the total received RF power at the UAV-RIS unit. If constraints such as battery capacity or IoT node QoS requirements are violated, the reward is penalized, guiding the UAV-RIS system toward an optimal and sustainable operational strategy.
\end{itemize}

To efficiently address this high-dimensional optimization problem, we propose the DDPG-EH algorithm, an enhanced version of Twin Delayed DDPG (TD3) \cite{fujimoto2018}. This approach mitigates learning challenges such as overestimation bias, exploration inefficiencies, and policy instability, ensuring robust decision-making in dynamic UAV-mounted RIS environments.

%*********************************************************************
\subsection{Energy-Efficient DRL for UAV-Mounted RIS Optimization}
\begin{figure*}[t]
\centering
\centerline{\includegraphics[width=2\columnwidth]{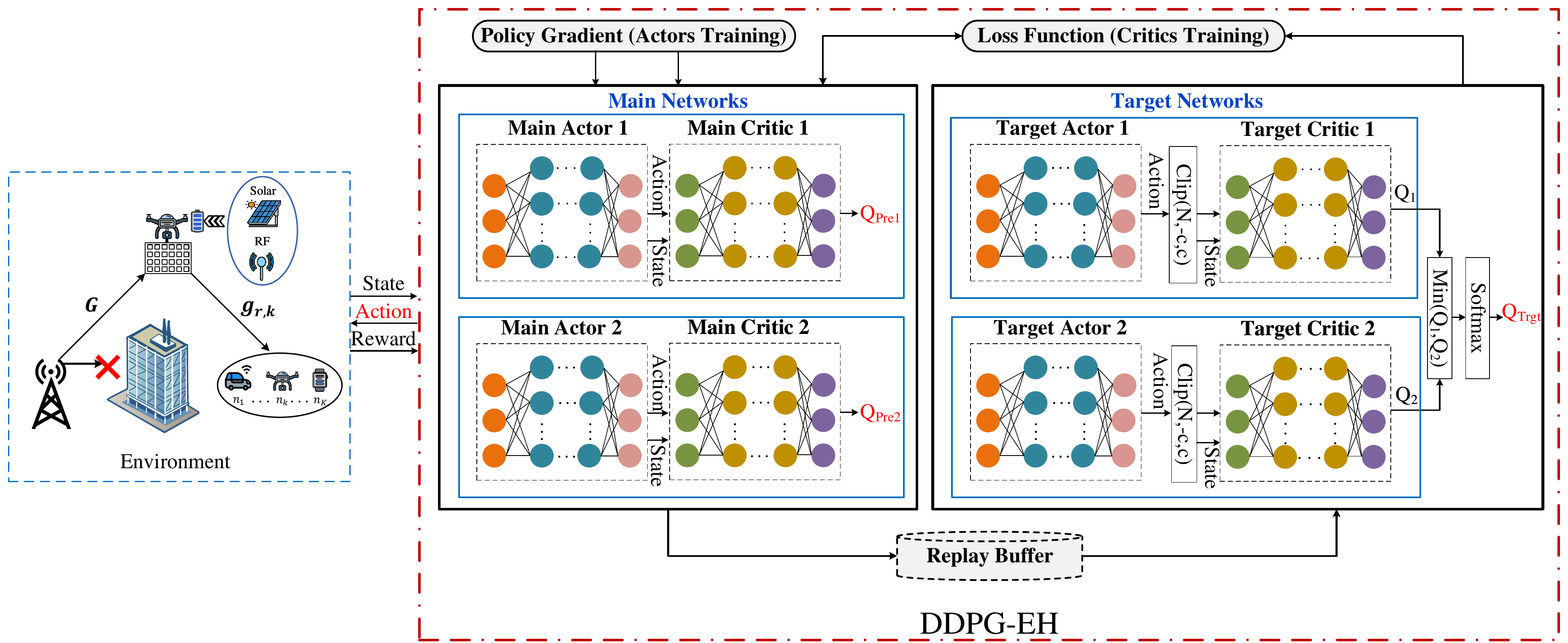}}
% \captionsetup{justification=centering}
\caption{Proposed DDPG-EH framework for solving UAV-mounted RIS optimization problem.}
\label{fig_DDPG}
\end{figure*}

%*********************************************************************
The proposed DDPG-based DRL framework optimizes UAV-mounted RIS networks by enhancing EH efficiency and adapting to environmental dynamics in real time. Through an advanced actor-critic architecture, the framework ensures stable and efficient resource allocation while incorporating action clipping and softmax-adjusted double Q-learning to mitigate overestimation and underestimation biases. In Fig. \ref{fig_DDPG}, the DDPG-EH framework consists of two independent actor-critic pairs with their corresponding target networks. The actor networks generate actions based on the observed system state, adjusting the aforementioned action parameters dynamically. The critic networks evaluate these actions by estimating their expected Q-values, guiding the optimization process. Target networks provide stabilization by computing target Q-values, ensuring smoother learning convergence.

% **********************************************************************
% *********************************************************************

To improve training efficiency, past interactions between the agent and the environment are stored in an experience replay buffer, breaking the correlation between consecutive updates and stabilizing learning. During each iteration, a mini-batch of transitions is sampled from the buffer, containing state, action, reward, next state, and terminal status tuples. This ensures learning from diverse experiences, enhancing generalization, and preventing overfitting. Once a batch of transitions is retrieved, the target actor networks generate exploratory actions incorporating controlled noise and action clipping to maintain exploration within valid bounds. The target critics evaluate these actions, selecting the minimum Q-value to counteract overestimation bias. A softmax operator further refines this estimation, balancing exploration and exploitation. The refined target Q-value ($Q_{\text{Trgt}}$ in Fig. \ref{fig_DDPG}) is then used within the Bellman equation to provide a stable reference for updating the main critics, ensuring robust policy optimization.

The training process follows a structured sequence to ensure stable convergence and policy refinement. The computed target Q-value ($Q_{\text{Trgt}}$) serves as a reference for updating the critic networks, where the predicted Q-value estimates, $Q_{\text{Pre1}}$ and $Q_{\text{Pre2}}$, are used to minimize the loss function and enhance value estimation accuracy. Following critic updates, the main actor networks undergo policy optimization, leveraging critic gradient feedback to refine action selection. This iterative learning cycle enables UAV-mounted RIS to dynamically adapt to fluctuating energy and communication conditions. The final policy selection follows an adaptive approach, ensuring that the most effective trained actor-critic pair is employed during execution. Upon training completion, both actor networks propose candidate actions, evaluated by their respective critics. The actor yielding the highest Q-value is selected, guaranteeing decision-making aligns with the most optimized policy. This dynamic selection mechanism enhances UAV-RIS responsiveness to variations in energy availability, channel dynamics, and mobility constraints, facilitating intelligent real-time adjustments.

By integrating clipped double Q-learning, softmax-adjusted Q-value estimation, and a structured actor-critic training framework, the proposed DDPG-EH model achieves robust and energy-efficient optimization of UAV-mounted RIS networks. The systematic learning process strikes a balance between maximizing EH efficiency and ensuring reliable communication, positioning the framework as a scalable and adaptive solution for next-generation energy-aware UAV-assisted networks.

% *******************************************************
\section{Results and Discussion}
% *******************************************************
% \section{Simulation Results and Performance Evaluation}

The simulation evaluates a multi-IoT node scenario with \( K = 3 \) mobile IoT nodes, where the UAV-mounted RIS operates under one of the three EH protocols, TS, PS, and the proposed hybrid ES-TS-PS, explained in Fig. \ref{fig_SM}(b). Additionally, the UAV-RIS unit harvests solar energy based on a discrete-time EH model, with energy arrival times governed by a Poisson distribution at an average rate of \( \lambda \) J/s \cite{salim-2024}. To adapt to IoT node mobility, the UAV employs the K-means clustering algorithm at each time slot, minimizing the distance between the UAV and the IoT nodes' cluster centroid to optimize trajectory \cite{dudik-2015}. The BS is equipped with \( Z = 8 \) antennas, while the UAV-mounted RIS consists of \( L = 16 \) reflective elements. Each IoT node ensures a minimum QoS of 70 Mbps. Furthermore, we consider the imperfect CSI estimated using the minimum mean squared error approach with an error of $\zeta=0.01$ and hardware impairment as an interference component of $\phi=0.08$.

\begin{figure}[th]
\centering
\centerline{\includegraphics[width=\columnwidth]{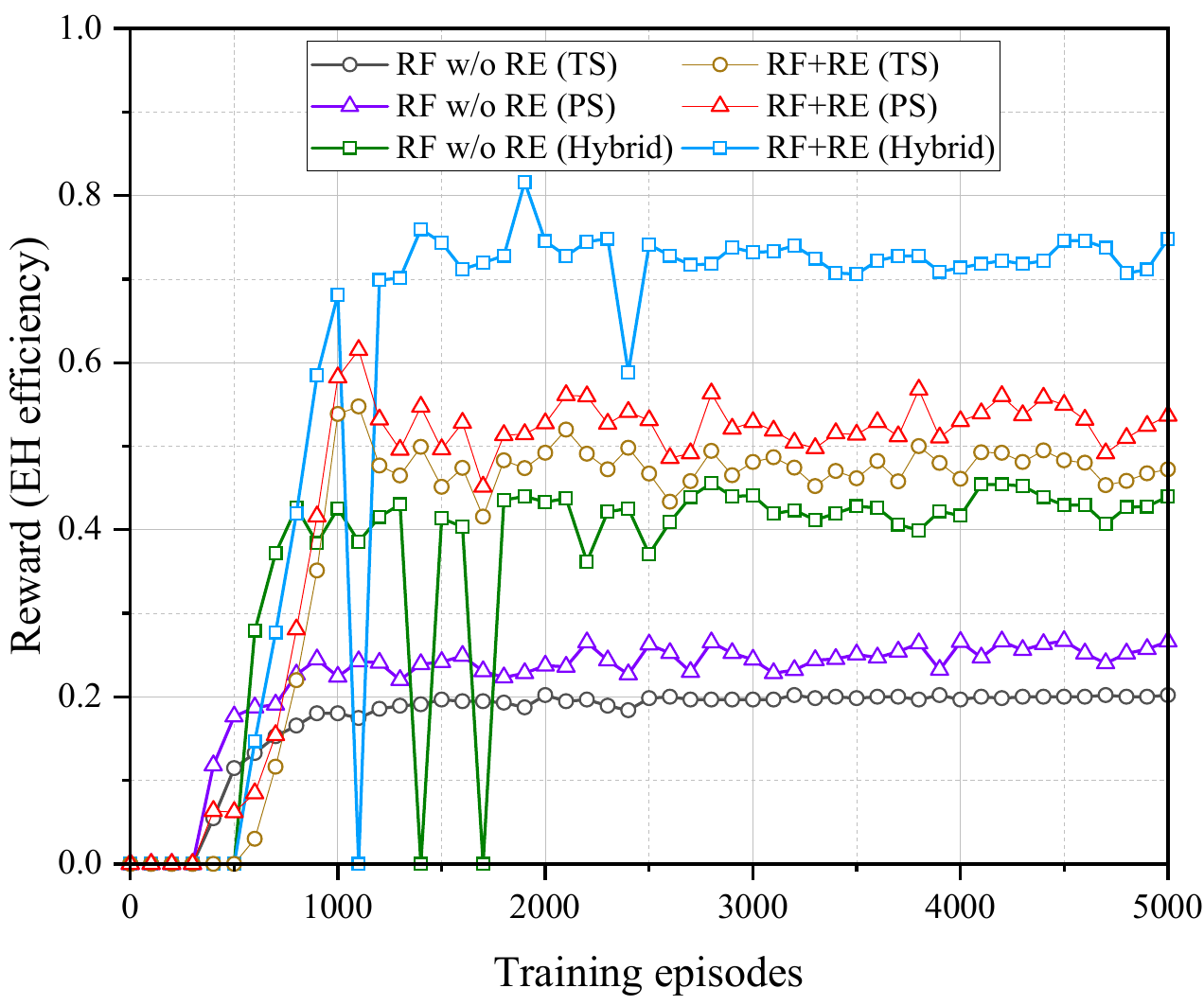}}
% \captionsetup{justification=centering}
\caption{The reward as EH efficiency vs the number of training episodes.}
\label{fig_rewards}
\end{figure}

\begin{figure}[th]
\centering
\centerline{\includegraphics[width=\columnwidth]{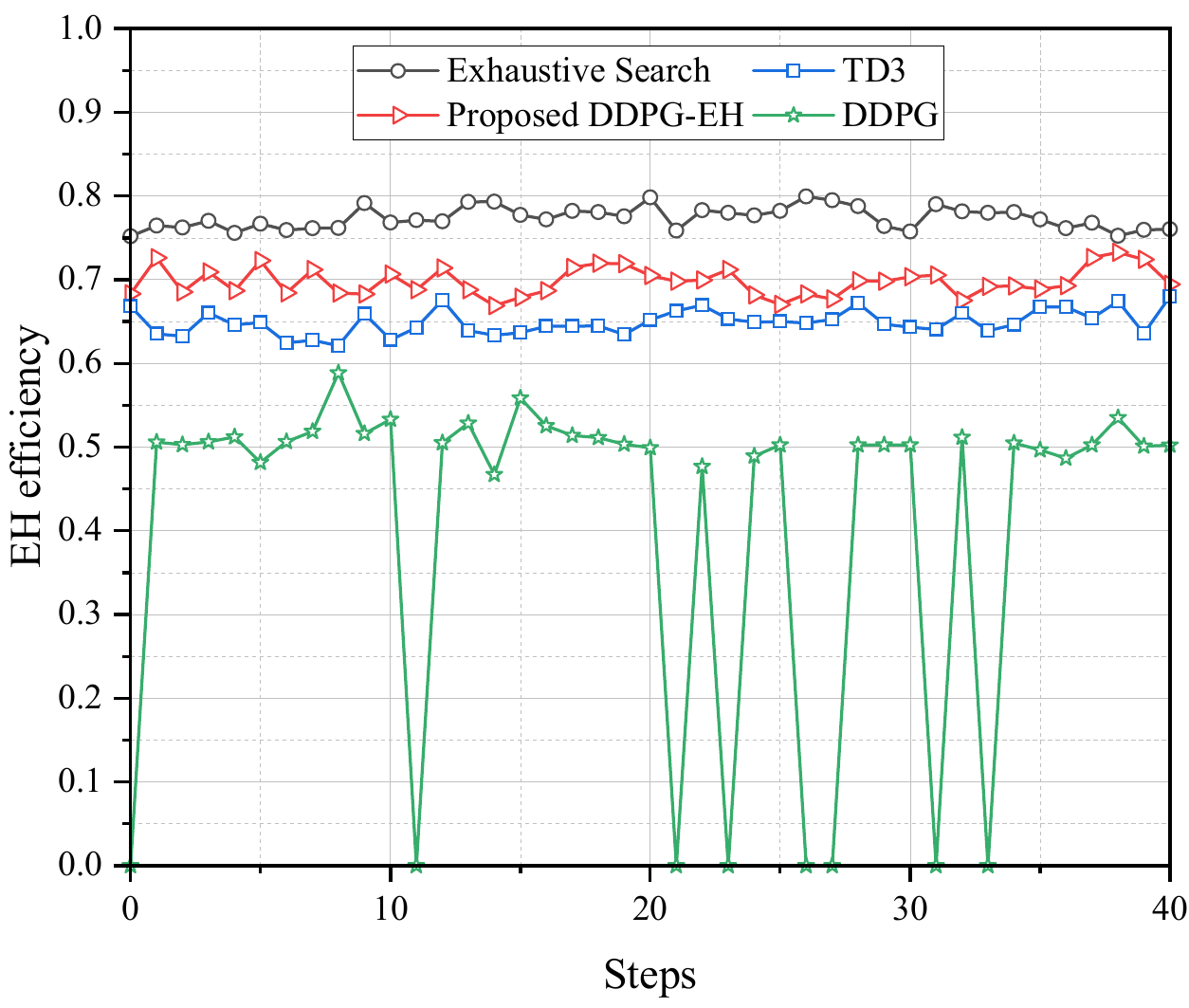}}
% \captionsetup{justification=centering}
\caption{EH efficiency per testing step.}
\label{fig_EH_Eff}
\end{figure}

\begin{figure}[th]
\centering
\centerline{\includegraphics[width=\columnwidth]{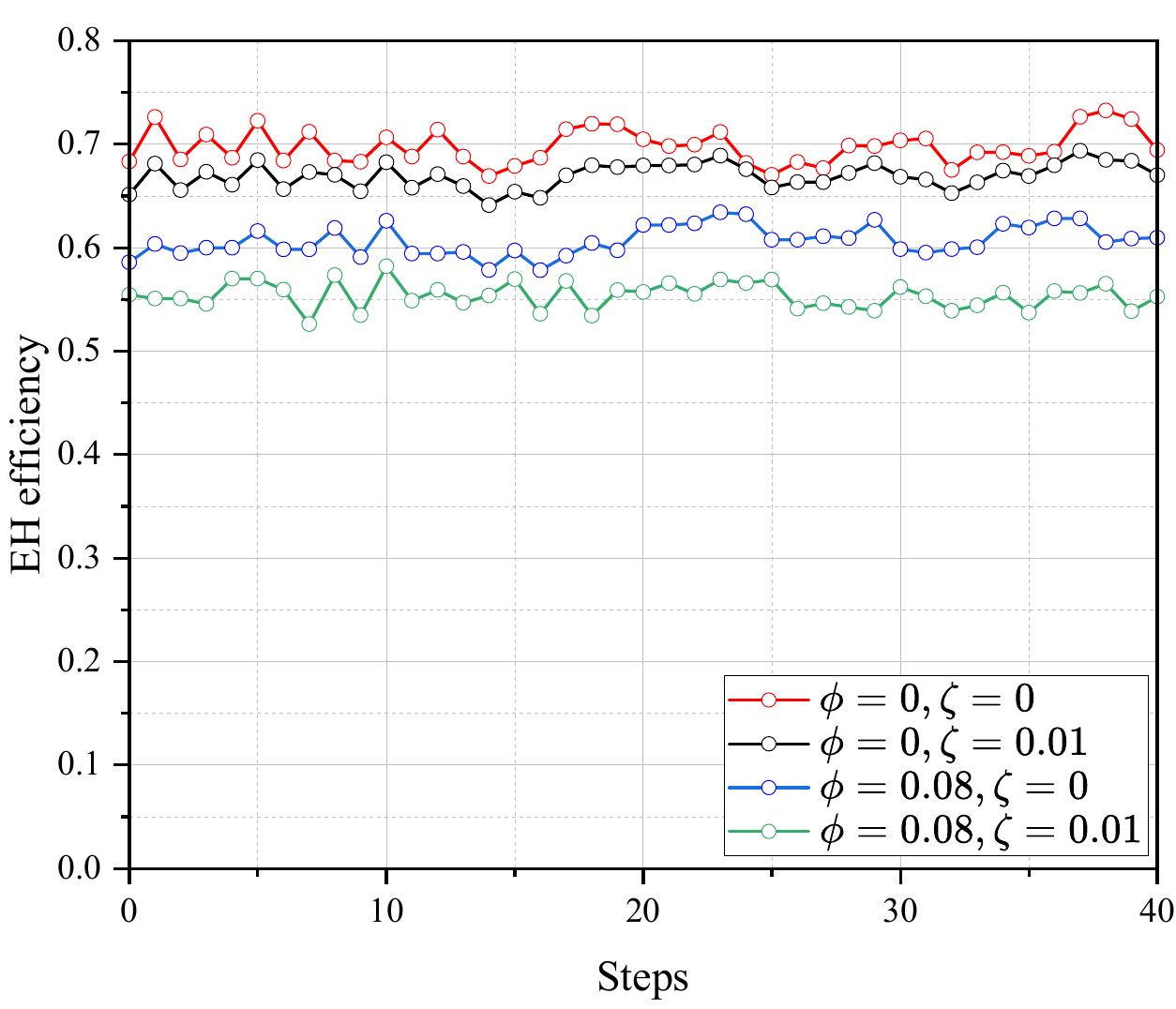}}
% \captionsetup{justification=centering}
\caption{DDPG-EH performance under impairments.}
\label{fig_impair}
\end{figure}

\begin{itemize}
    \item \textbf{Reward Performance and EH Protocols}: Fig. \ref{fig_rewards} presents the reward convergence of different EH protocols (TS, PS, and hybrid ES-TS-PS) with and without (w/o) RE using the proposed DDPG-EH algorithm. Among the protocols, TS exhibits the lowest performance, while PS shows a slight improvement. The hybrid approach achieves the highest EH efficiency, demonstrating its advantage in optimizing EH. Additionally, incorporating RE as a complementary source significantly enhances EH efficiency across all protocols. However, the hybrid approach experiences greater fluctuations in reward convergence and more instances of reward collapse due to the increased complexity of the optimization problem.
    \item \textbf{EH Efficiency Performance of DDPG-EH vs. Benchmarks}: Fig. \ref{fig_EH_Eff} illustrates the EH efficiency performance of the proposed DDPG-EH algorithm in comparison with exhaustive search, TD3, and DDPG approaches over testing steps. The results indicate that exhaustive search achieves the highest performance by identifying the global optimum but suffers from high computational complexity and lacks adaptability to network dynamics. In contrast, the DDPG-EH algorithm achieves performance close to exhaustive search while maintaining lower complexity, offering a more intelligent and efficient solution. Additionally, DDPG-EH outperforms TD3 due to action clipping and softmax-weighted Q-value estimation, effectively mitigating underestimation and overestimation issues. These enhancements improve stability and exploration-exploitation balance, resulting in superior EH efficiency. Conversely, the DDPG algorithm demonstrates the lowest performance due to overestimation bias and unstable learning, leading to convergence issues and suboptimal EH decisions. 
    % Table \ref{tab:general_comparison} summarizes the comparison, justifying the selection of DDPG-EH for its balanced trade-off between learning efficiency, stability, and EH performance.
    \item \textbf{Impairments Effect}: The impact of hardware impairments and CSI imperfections on the EE-DDPG algorithm is shown in Fig. \ref{fig_impair}. The algorithm achieves optimal EH under ideal conditions but experiences a slight efficiency drop with CSI imperfections when $\zeta=0.01$, reflecting its sensitivity to estimation errors. hardware impairments with $\phi=0.08$ cause a more significant EH reduction, highlighting resource management challenges in non-ideal hardware. When both impairments are present, degradation worsens, impacting IoT node QoS, which increases energy demands to maintain the minimum IoT node data rate.    
\end{itemize}

% \begin{table*}[h]
% \centering
% \caption{Comparison of Optimization Approaches in Reinforcement Learning}
% \label{tab:general_comparison}
% \resizebox{\textwidth}{!}{ % Automatically scale table to fit within page width
% \begin{tabular}{|p{3cm}|p{2.5cm}|p{2.5cm}|p{3cm}|p{3cm}|p{4cm}|}  % Adjusted column widths
% \hline
% \textbf{Approach} & \textbf{Computational Complexity} & \textbf{Convergence Speed} & \textbf{Bias Handling} & \textbf{Exploration-Exploitation} & \textbf{Applicability} \\\hline
% Exhaustive Search & Very high & Slow & No bias (Optimal) & N/A & Theoretical benchmark but impractical for large-scale problems \\\hline
% DDPG & Moderate & Fast & Prone to overestimation bias & Poor tradeoff, leading to instability & Suitable for continuous control but requires extensive tuning \\\hline
% TD3 & Moderate & Faster than DDPG & Reduces overestimation but prone to underestimation & Conservative updates can slow adaptation & More stable than DDPG, preferred for continuous action spaces \\\hline
% Proposed DDPG-EH & Moderate & Fastest & Balanced overestimation and underestimation due to action clipping and softmax & Improved tradeoff for faster learning & Enhanced stability and efficiency in dynamic environments \\\hline
% \end{tabular}
% } % End resizebox
% \end{table*}

% **********************************************************************
% *********************************************************************

% **********************************************************************
% *********************************************************************
% **********************************************************************
% *********************************************************************
% *****************************************************
\section{Future Directions in UAV-Mounted RIS Networks}
% *****************************************************
This section presents and discusses some potential future research directions.

\subsection{Quantum-Assisted Optimization}
% ******************************************************
Optimizing UAV trajectory, RIS phase shifts, and EH scheduling in UAV-mounted RIS networks is computationally intensive, often leading to suboptimal energy utilization and reduced endurance. Traditional methods struggle with high-dimensional decision spaces, limiting real-time adaptability. However, quantum-assisted optimization can significantly accelerate learning and decision-making, enabling UAVs to rapidly identify optimal EH and trajectory strategies. Future research can explore quantum-enhanced DRL for efficient resource allocation and quantum key distribution to secure EH transactions, ensuring energy-efficient UAV operation in dynamic 6G-enabled IoT networks.

% ******************************************************
\subsection{Multi-UAV Cooperative RIS Systems with Federated DRL}
% ******************************************************
In a large-scale IoT deployment, multiple UAV-RIS systems will be used to work cooperatively. In such scenarios, federated reinforcement learning can be utilized to train agents without sharing raw data which can reduce latency in policy updates and further improve the network security.
% ******************************************************
\subsection{Blockchain-Enabled Security and Trust}
% ******************************************************
Security and trust in UAV-RIS-aided IoT networks remain critical, particularly in multi-UAV deployments where unauthorized access and cyber threats can disrupt EH and resource allocation. Without robust security, UAVs face risks such as energy theft, inefficient power distribution, and compromised network coordination, ultimately degrading EH efficiency and UAV endurance. Blockchain technology can enable decentralized authentication, secure energy trading, and interference management. Future research can explore blockchain-based smart contracts for autonomous EH transactions and tamper-proof ledgers to track UAV energy consumption, flight paths, and RIS configurations. Securing EH processes through blockchain can also optimize energy distribution, prevent power loss, and enhance UAV longevity in dynamic aerial networks.

% ******************************************************
% \subsection{AI-Driven Adaptation and UAV Swarm Intelligence}
% ******************************************************
% Coordinating multi-UAV RIS networks is challenging due to the need for efficient EH utilization and optimized coverage. Isolated UAV operations often lead to energy wastage and suboptimal resource distribution. Swarm-based UAV intelligence enables UAVs to collaborate on beamforming, trajectory planning, and energy allocation, maximizing EH efficiency. Future research can explore federated learning to allow UAVs to train AI models for energy-aware decision-making while minimizing power-intensive data exchange. Additionally, DRL-based adaptive scheduling can dynamically adjust EH priorities based on mobility, interference, and IoT node demands, enhancing energy efficiency and extending UAV operation time in large-scale RIS-assisted networks.

% ******************************************************
\subsection{ML-based Channel Estimation Algorithms}
% ******************************************************
Accurate channel estimation is crucial for optimizing EH efficiency in UAV-RIS networks, directly influencing RIS phase shifts, power allocation, and beamforming accuracy. However, imperfect CSI due to UAV mobility and Doppler shifts can hinder RIS elements from efficiently harvesting energy, reducing overall power availability. Frequent UAV movement alters LoS and NLoS conditions, making EH optimization challenging. ML-driven adaptive channel estimation can enhance CSI accuracy, enabling real-time EH scheduling and RIS reconfiguration, significantly improving energy efficiency and UAV endurance.
% ******************************************************
\subsection{Millimeter-Wave Beamforming}
% ******************************************************

Millimeter-wave communication at 28 GHz and beyond supports ultra-high data rates but suffers from severe path loss and signal blockages, which are exacerbated at sub-terahertz and terahertz frequencies in 6G-enabled IoT networks. While the proposed EH framework in this work improves UAV endurance under current operating conditions, supporting mmWave UAV-RIS communication introduces additional challenges due to higher power demands and increased beam misalignment caused by UAV mobility. In this context, AI-driven dynamic beamforming is essential to optimize beam tracking and alignment, minimizing energy consumption and compensating for mobility-induced fluctuations. Furthermore, predictive UAV trajectory planning can improve EH efficiency by maintaining optimal LoS paths, reducing energy waste, and supporting long-duration operations. Future research should explore EH-aware mmWave beamforming to enable sustainable and energy-efficient UAV-RIS communication in high-frequency 6G-enabled IoT environments.

% ******************************************************
\section{Conclusion}
% ******************************************************
This paper proposed a hybrid EH framework and a DRL-based optimization strategy for more energy-efficient 6G-enabled IoT networks. We outlined the system architecture and EH operating protocols, introducing the proposed hybrid ES-TS-PS EH strategy, which integrates RF and solar energy to sustain UAV operation. Key application scenarios were presented highlighting the potential of UAV-mounted RIS for wireless coverage extension, disaster recovery, and smart city infrastructure. Furthermore, a comprehensive case study was conducted to evaluate the performance of the proposed DRL-based system in optimizing UAV trajectory, RIS phase shifts, and EH scheduling. The results confirmed significant improvements in EH efficiency and system performance over several baseline methods. Furthermore, we explored key challenges and potential research directions to fully harness the capabilities of this technology for emerging applications.

\bibliographystyle{IEEEtran}
\bibliography{Main}

\begin{thebibliography}{10}
\providecommand{\url}[1]{#1}
\csname url@samestyle\endcsname
\providecommand{\newblock}{\relax}
\providecommand{\bibinfo}[2]{#2}
\providecommand{\BIBentrySTDinterwordspacing}{\spaceskip=0pt\relax}
\providecommand{\BIBentryALTinterwordstretchfactor}{4}
\providecommand{\BIBentryALTinterwordspacing}{\spaceskip=\fontdimen2\font plus
\BIBentryALTinterwordstretchfactor\fontdimen3\font minus \fontdimen4\font\relax}
\providecommand{\BIBforeignlanguage}[2]{{%
\expandafter\ifx\csname l@#1\endcsname\relax
\typeout{** WARNING: IEEEtran.bst: No hyphenation pattern has been}%
\typeout{** loaded for the language `#1'. Using the pattern for}%
\typeout{** the default language instead.}%
\else
\language=\csname l@#1\endcsname
\fi
#2}}
\providecommand{\BIBdecl}{\relax}
\BIBdecl

\bibitem{wang-2023}
C.-X. Wang, X.~You, X.~Gao, X.~Zhu, Z.~Li, C.~Zhang, H.~Wang, Y.~Huang, Y.~Chen, H.~Haas, J.~S. Thompson, E.~G. Larsson, M.~Di~Renzo, W.~Tong, P.~Zhu, X.~Shen, H.~V. Poor, and L.~Hanzo, ``{On the road to 6G: visions, requirements, key technologies, and testbeds},'' \emph{IEEE Commun. Surveys Tuts.}, vol.~25, no.~2, pp. 905--974, Jan. 2023.

\bibitem{keytech-ji-2021}
B.~Ji, Y.~Han, S.~Liu, F.~Tao, G.~Zhang, Z.~Fu, and C.~Li, ``{Several key technologies for 6G: challenges and opportunities},'' \emph{IEEE Commun. Stand. Mag.}, vol.~5, no.~2, pp. 44--51, June 2021.

\bibitem{shi-2022}
W.~Shi, W.~Xu, X.~You, C.~Zhao, and K.~Wei, ``{Intelligent reflection enabling technologies for integrated and green Internet-of-Everything beyond 5G: communication, sensing, and security},'' \emph{IEEE Wirel. Commun.}, vol.~30, no.~2, pp. 147--154, May 2022.

\bibitem{zhou-2023}
H.~Zhou, M.~Erol-Kantarci, Y.~Liu, and H.~V. Poor, ``{A survey on model-based, heuristic, and machine learning optimization approaches in RIS-aided wireless networks},'' \emph{IEEE Commun. Surveys Tuts.}, vol.~26, no.~2, pp. 781--823, Dec. 2023.

\bibitem{deng-2024}
M.~Deng, M.~Ahmed, A.~Wahid, A.~A. Soofi, W.~U. Khan, F.~Xu, M.~Asif, and Z.~Han, ``{Reconfigurable intelligent surfaces enabled vehicular communications: a comprehensive survey of recent advances and future challenges},'' \emph{IEEE Trans. Intell. Veh.}, pp. 1--28, 1 2024.

\bibitem{ris-satinoma-liu-2024}
R.~Liu, K.~Guo, X.~Li, K.~Dev, S.~A. Khowaja, T.~A. Tsiftsis, and H.~Song, ``{RIS-empowered satellite-aerial-terrestrial networks with PD-NOMA},'' \emph{IEEE Commun. Surv. Tutor.}, vol.~26, no.~4, pp. 2258--2289, Jan. 2024.

\bibitem{li-2020}
S.~Li, B.~Duo, X.~Yuan, Y.-C. Liang, and M.~Di~Renzo, ``{Reconfigurable intelligent surface assisted UAV communication: joint trajectory design and passive beamforming},'' \emph{IEEE Wirel. Commun. Lett.}, vol.~9, no.~5, pp. 716--720, Jan. 2020.

\bibitem{ahmad-2022}
I.~Ahmad, R.~Narmeen, Z.~Becvar, and I.~Guvenc, ``{Machine learning-based beamforming for unmanned aerial vehicles equipped with reconfigurable intelligent surfaces},'' \emph{IEEE Wirel. Commun.}, vol.~29, no.~4, pp. 32--38, Aug. 2022.

\bibitem{liu-2018}
J.~Liu, Y.~Shi, Z.~M. Fadlullah, and N.~Kato, ``{Space-air-ground integrated network: a survey},'' \emph{IEEE Commun. Surveys Tuts.}, vol.~20, no.~4, pp. 2714--2741, Jan. 2018.

\bibitem{ghosh-2024}
\BIBentryALTinterwordspacing
S.~Ghosh, A.~Bhowmick, S.~D. Roy, and S.~Kundu, ``{UAV-RIS enabled NOMA network for disaster management with hardware impairments},'' \emph{IEEE Trans. Aerosp. Electron. Syst.}, pp. 1--13, Jan. 2024. [Online]. Available: \url{https://doi.org/10.1109/taes.2024.3421177}
\BIBentrySTDinterwordspacing

\bibitem{peng-2023}
H.~Peng and L.-C. Wang, ``{Energy harvesting reconfigurable intelligent surface for UAV based on robust deep reinforcement learning},'' \emph{IEEE Trans. Wirel. Commun.}, vol.~22, no.~10, pp. 6826--6838, Feb. 2023.

\bibitem{salim-2022}
M.~M. Salim, H.~A. Elsayed, M.~A. Elaziz, M.~M. Fouda, and M.~S. Abdalzaher, ``{An optimal balanced energy harvesting algorithm for maximizing Two-Way relaying D2D communication data rate},'' \emph{IEEE Access}, vol.~10, pp. 114\,178--114\,191, Jan. 2022.

\bibitem{fujimoto2018}
S.~Fujimoto, H.~Hoof, and D.~Meger, ``{Addressing function approximation error in actor-critic methods},'' in \emph{Proc. 35th Int. Conf. Mach. Learn. (ICML)}.\hskip 1em plus 0.5em minus 0.4em\relax PMLR, Jul. 2018, pp. 1587--1596.

\bibitem{salim-2024}
M.~M. Salim, S.~I. Al-Dharrab, D.~B. Da~Costa, and A.~H. Muqaibel, ``{Rate-energy optimization for hybrid-powered full-duplex relays in cognitive C-NOMA with impairments},'' \emph{IEEE Open J. Commun. Soc.}, p.~1, Jan. 2024.

\bibitem{dudik-2015}
J.~M. Dudik, A.~Kurosu, J.~L. Coyle, and E.~Sejdić, ``{A comparative analysis of DBSCAN, K-means, and quadratic variation algorithms for automatic identification of swallows from swallowing accelerometry signals},'' \emph{Comput. Biol. Med.}, vol.~59, pp. 10--18, Apr. 2015.

\end{thebibliography}

% \newpage

 

% \end{IEEEbiography}



% \vfill

\end{document}